\documentclass{elsart}

\usepackage{graphicx}
\usepackage[english]{babel}
\usepackage{latexsym}

\begin{document} 

\begin{frontmatter}

\title{Supernovae and Radio Galaxies Probing Gurzadyan-Xue Cosmological Models with Dark Energy}

\author{G.V. Vereshchagin \and G. Yegorian}
\ead{veresh@icra.it}
\ead{gegham@icra.it}

\address{ICRANet, P.le della Repubblica 10, I–65100 Pescara, Italy and \\
ICRA, Dip. Fisica, Univ. ``La Sapienza'', P.le A. Moro 5, I–00185 Rome, Italy}


\begin{abstract}
In the previous papers we derived cosmological equations for models with Gurzadyan-Xue dark energy, have performed their qualitative analysis and, particularly, have revealed a remarkable hidden invariance in the models with respect
to the separatrix $\Omega_{sep}$ in their phase portraits. Now, new analytic solutions for these models are obtained, showing additional symmetries at various curvatures. The likelihood analysis with supernovae and radio galaxies data and their characteristics (age, the deceleration parameter), again demonstrate the crucial role of the $\Omega_{sep}$ for all models, in spite of the diversity of both, the initial equations and their solutions. 
\end{abstract}

\end{frontmatter}

\section{Introduction}

The formula for dark energy, derived by Gurzadyan and Xue \cite{GX} represents a scaling between the speed of light $c$, gravitational constant $G$ and scale factor $a$ of the Universe
\begin{eqnarray}
\label{rhoLambda}
\rho_{GX}=\frac{\pi}{4}\,\frac{\hbar c}{L_p^2}\,\frac{1}{a^2}=\frac{\pi}{4}\,\frac{c^4}{G}\,\frac{1}{a^2},
\end{eqnarray}
where $\hbar$ is Planck's constant, $L_p$ is Planck's length. The formula is obtained via estimation of the energy
of the relevant vacuum fluctuations and remarkably fits the SN data for the current value of the dark energy. 

Certain observational aspects of the GX-formula (1) are discussed in \cite{DG}. This formula can be interpreted as indication on possible fundamental constants variation with cosmic time.  In particular, in order to keep interpretation of the vacuum energy as the cosmological term one has to suppose that the speed of light is proportional to the scale factor as studied in \cite{Ver06}. In other model gravitational constant changes with time (see similar model following from holographic approach \cite{Horvat}).

Based on the scaling (\ref{rhoLambda}) we have proposed a set of cosmological models \cite{Ver06} and performed their general analysis \cite{Ver06a} using phase portrait technique. Although cosmological equations describing each model are very different, some interesting similarities were found.

In particular, qualitative analysis shows that there is a separatrix in the phase space of dynamical variables and it corresponds to particular solution of cosmological equations with usual matter density parameter equal to 2/3 for almost all cases. In this way all solutions with $\Omega_m>\Omega_{sep}$ contain Friedmannian singularity in the past and have Friedmannian asymptotic in the future as well, while all solutions with $\Omega_m<\Omega_{sep}$ begin with some positive and large scale factor and zero density.

In this Letter we find new analytic solutions of the models and compare them with observations on supernovae and radio galaxies. We then perform likelihood analysis and provide best-fit values of the density parameter for all models and also calculate such important quantities as the age and deceleration parameters.

\section{GX models}

In general, with variation of the gravitational constant, speed of light and cosmological constant $\Lambda$ with time we have \cite{Ver06b} 
\begin{eqnarray}
\label{fe1}
H^2+\frac{k c^2}{a^2}-\frac{\Lambda}{3}=\frac{8\pi G}{3}\mu,
\end{eqnarray}
\begin{eqnarray}
\label{mue}
\dot\mu+3H\left(\mu+\frac{p}{c^2}\right)=-\mu\frac{\dot G}{G}-\frac{\dot\Lambda}{8\pi G}+\frac{3}{4\pi G}\,\frac{\dot c}{c}\left(H^2+\frac{k c^2}{a^2}\right),
\end{eqnarray}
where $k=-1,0,1$ is curvature parameter, $\mu$ is matter density, $p$ is pressure, $H$ is Hubble parameter, $a$ is the scale factor of the Universe.
Introducing instead of $\frac{d}{dt}$ derivatives with respect to scale factor $\frac{d}{da}$, with $p=0$ we arrive to
\begin{eqnarray}
	\frac{d\mu}{da}+3\frac{\mu}{a}+\mu\frac{dG}{Gda}+\frac{1}{8\pi G}\,\frac{d\Lambda}{da}-\frac{1}{4\pi G}\,\frac{dc}{cda}\left(8\pi G\mu+\Lambda\right)=0.
\end{eqnarray}

Then, as in usual Friedmannian cosmology we introduce density parameter
\begin{eqnarray}
	\Omega_m=\frac{8\pi G}{3 H_0^2}\mu_0,
	\label{Om}
\end{eqnarray}
where $H_0$ is Hubble constant.

{\bf Model I.}
Neither the speed of light nor the gravitational constant vary with time, but $\Lambda\propto a^{-2}$. Cosmological equations are

\begin{eqnarray}
	H^2+\left(k-\frac{2\pi^2}{3}\right)\frac{c^2}{a^2}=\frac{8\pi G}{3}\mu, \quad
	\dot\mu+3H\mu=\frac{\pi}{2G}\left(\frac{c}{a}\right)^2 H.
\end{eqnarray}

If one neglects the source term in the continuity equation (that is possible for sufficiently large scale factor), this system describes the usual Friedmannian cosmology with negative curvature.

There is analytic solution for density which reads
\begin{eqnarray}
	\mu(a)=\mu_0\left(\frac{a_0}{a}\right)^3+\frac{\pi}{2G}\,\frac{c^2}{a^2}\left(1-\frac{a_0}{a}\right),
\end{eqnarray}
where $\mu(a_0)=\mu_0$. Taking the limit of this expression with $a\rightarrow 0$ one can find the value of $\Omega_m$ using (\ref{Om}) for separatrix \cite{Ver06a} which turns out to be
\begin{eqnarray}
	\Omega{'}_{sep}=\frac{4\pi^2}{3(2\pi^2-k)},
	\label{sepprime}
\end{eqnarray}
having values $\Omega_{sep}\approx0.6345$ for $k=-1$, $\Omega_{sep}=2/3$ for $k=0$ and $\Omega_{sep}\approx0.7022$ for $k=1$.
Solution for the scale factor is
\begin{equation}
	\begin{array}{l}
	\pm A^\frac{3}{2}t=\sqrt{a A_1(a A_1+B_1)}-\sqrt{a_0 A_1(a_0 A_1+B_1)}+ \\ \displaystyle{\quad\quad\quad\quad\quad\quad\quad\quad\quad\quad\quad\quad\quad +B_1\log\left(\frac{\sqrt{a_0 A_1}+\sqrt{a_0 A_1+B_1}}{\sqrt{a A_1}+\sqrt{a A_1+B_1}}\right)},
	\label{deq1}
	\end{array}
\end{equation}
where
\begin{eqnarray}
	A_1=c^2(2\pi^2-k), \quad
	B_1=\frac{4\pi}{3}a_0(2G\mu_0 a_0^2-\pi c^2).
\end{eqnarray}
This solution reduces to a simple form for separatrix (\ref{sepprime}) with the corresponding condition $B_1=0$
\begin{eqnarray}
	a=a_0+\sqrt A t.
	\label{ballistic}
\end{eqnarray}

{\bf Model II.}
Speed of light changes with time, $c=\left(\frac{\Lambda}{2\pi^2}\right)^{1/2}a$, the gravitational constant does not change and  $\Lambda$=const. Cosmological equations reduce to the following system

\begin{eqnarray}
	H^2-\frac{\Lambda}{3}\left(1-\frac{3k}{2\pi^2}\right)=\frac{8\pi G}{3}\mu, \quad
	\dot\mu+3H\mu=\frac{3H}{4\pi G}\left(H^2+\frac{k\Lambda}{2\pi^2}\right),
\end{eqnarray}
If one neglects the source in the last equation (that is again possible for large $a$), which consists of two terms, proportional to $H^3$ and $H\Lambda$ correspondingly, we come to the usual Friedmann-Lema\^itre cosmology with cosmological constant $\Lambda'$.

There is complete analytical solution for this model
\begin{eqnarray}
	\mu(a)=\mu_0\frac{a_0}{a}+\frac{\Lambda}{4\pi G}\left(1-\frac{a_0}{a}\right),
\end{eqnarray}
\begin{equation}
	\begin{array}{l}
	\displaystyle{a(t)=-\frac{A_2}{2B_2}+\frac{A+2a_0 B_2}{2B_2}\cosh\left(t\sqrt B_2\right)\pm} \\ \displaystyle{\quad\quad\quad\quad\quad\quad\quad\quad\quad\quad\quad\quad\quad\quad\quad \pm\sqrt{\frac{a_0}{B_2}\left(A_2+a_0 B_2\right)}\sinh\left(t\sqrt B_2\right)},
	\label{sol2}
	\end{array}
\end{equation}
where
\begin{eqnarray}
	A_2=\frac{2}{3}a_0\left(4\pi G\mu_0-\Lambda\right), \quad
	B_2=\Lambda\left(1-\frac{k}{2\pi^2}\right).
\end{eqnarray}
Separatrix for this model is given by
\begin{eqnarray}
	\Omega_{sep}=\frac{2}{3}
	\label{sep}
\end{eqnarray}
independently on spatial curvature.

Solution (\ref{sol2}) has particularly simple form with $\Omega_m=\Omega_{sep}$ which is
\begin{eqnarray}
	a=a_0\exp\left\{\left(1-\frac{k}{2\pi^2}\right)\Lambda t\right\}.
	\label{sol2sep}
\end{eqnarray}

{\bf Model III.}
The gravitational constant changes, $G=\frac{\pi}{4\mu_{GX}}\left(\frac{c}{a}\right)^2$, while the speed of light does not and $\Lambda\propto a^{-2}$. Cosmological equations are

\begin{eqnarray}
H^2+\frac{k c^2}{a^2}=\frac{2\pi^2}{3}\left(\frac{c}{a}\right)^2\left(1+\frac{\mu}{\mu_{GX}}\right), \quad
	\dot\mu+3H\mu=2H\mu\left(1+\frac{\mu_{GX}}{\mu}\right),
\end{eqnarray}
where $\mu_{GX}\equiv\frac{\Lambda(a)}{8\pi G(a)}=$const.

These are very different from the usual Friedmann equations. Even if we neglect the source terms in the continuity equation, the first cosmological equation does not correspond to the first Friedmann one.

Again there is solution for density
\begin{eqnarray}
	\mu=\mu_0\frac{a_0}{a}+2\mu_{GX}\left(1-\frac{a_0}{a}\right).
\end{eqnarray}
Notice similarity with model II.

Solution for the scale factor looks exactly the same as (\ref{deq1}) with
\begin{eqnarray}
	A_3=A_1, \quad
	B_3=\frac{2\pi^2}{3\mu_{GX}}c^2 a_0(\mu_0-2\mu_{GX}).
\end{eqnarray}
The separatrix is again given by (\ref{sep}) with $B_3=0$. We have already presented in \cite{Ver06a} analytic solution for $\mu(t)$.

{\bf Model IV.}
The speed of light varies as $c=\left(\frac{4G\rho_{GX}}{\pi}\right)^{1/4}a^{1/2}$ and $G$=const. Here $\Lambda\propto a^{-1}$, and cosmological equations are

\begin{eqnarray}
	H^2=\frac{8\pi G}{3}\mu+\frac{\beta}{a}, \quad
	\dot\mu+3H\mu=\frac{3H}{8\pi G}\left(H^2+\frac{\beta}{a}\,\frac{2\pi^2+3k}{2\pi^2-3k}\right),
\end{eqnarray}
with $\beta=\frac{4\pi^2}{3}\left(G\rho_{GX}\right)^{1/2}\left(1-\frac{3k}{2\pi^2}\right)$, $\rho_{GX}\equiv\frac{\Lambda(a) c(a)^2}{8\pi G}=$const.

The source term in the continuity equation can be neglected for large $a$, but the first Friedmann equation contains the term $a^{-1}$ which is again very different from the standard cosmology.

There is complete solution for this model with
\begin{eqnarray}
	\mu=\mu_0\left(\frac{a_0}{a}\right)^2+\frac{1}{a}\sqrt\frac{\pi \rho_{GX}}{G}\left(1-\frac{a_0}{a}\right),
\end{eqnarray}
and
\begin{eqnarray}
	a=a_0\pm t\sqrt{A_4+a_0 B_4}+\frac{B_4}{4}t^2,
\end{eqnarray}
where
\begin{eqnarray}
	A_4=\frac{8\pi G}{3}a_0\left(\mu_0 a_0-\sqrt\frac{\pi\rho_{GX}}{G}\right), \quad
	B_4=2\sqrt\frac{G\rho_{GX}}{\pi}(2\pi^2-k).
\end{eqnarray}

This solution reduces to (\ref{ballistic}) for $B_4=0$ or, in other words, when $\Omega_m\rightarrow 1$.

\section{Age}

From (\ref{fe1}) in general case we have
\begin{eqnarray}
	t_0=\int_{a_i}^{a_0}\frac{da}{a}\left[\frac{8\pi G(a)}{3}\mu(a)+\frac{\Lambda(a)}{3}-\frac{kc(a)^2}{a^2}\right]^{-\frac{1}{2}},
\end{eqnarray}
where $a_i$ is either 0 (for $\Omega_m>\Omega_{sep}$) or the value of the scale factor for which the density vanishes (for $\Omega_m<\Omega_{sep}$).
There are explicit solutions for models II and IV, so the age can be computed directly, in particular for model IV the age with $\Omega_m\geq\Omega_{sep}$ is given by
\begin{eqnarray}
	t_0=\frac{2}{B_4}\left(\sqrt{A_4+a_0 B_4}-\sqrt A_4\right).
\end{eqnarray}
Coefficients $A$ and $B$ can be computed from (\ref{Om}) and $\Lambda(a_0)=3H_0^2(1-\Omega_m)$ with appropriate definitions of $\mu_{GX}$ and $\rho_{GX}$.

Results for all four models are represented in tab. \ref{table} for selected values of $\Omega_m$. We assumed $H_0=70 h$ Km/(s Mpc).
\begin{table*}[htp]
	\centering
\begin{tabular}
[c]{|c|c|c|c|c|c|c|c|c|c|}\hline
{\tiny k} & {\tiny $\Omega_{m}$} & \multicolumn{2}{|c|}{{\tiny I}} &
\multicolumn{2}{|c|}{{\tiny II}} & \multicolumn{2}{|c|}{{\tiny III}} &
\multicolumn{2}{|c|}{{\tiny IV}}\\\cline{3-10}
&  & {\tiny age,GYr} & {\tiny dec. par.} & {\tiny age,GYr} & {\tiny dec.
par.} & {\tiny age,GYr} & {\tiny dec. par.r} & {\tiny age,GYr} &
{\tiny dec. par.}\\\hline
& {\tiny 0.3} & {\tiny 3.72} & {\tiny -1.82(3.09)} &
{\tiny 3.44} & {\tiny -5.99(2.27)} & {\tiny 3.21} & {\tiny -1.96(2.28)}
& {\tiny 3.99} & {\tiny -3.88(2.33)}\\\cline{2-10}
{\tiny --1} & {\tiny $\Omega_{sep}$} & {\tiny 14.0} & {\tiny 0} & {\tiny $\infty$} &
{\tiny -4.00} & {\tiny 14.0} & {\tiny 0} & {\tiny 26.9} &
{\tiny -2.00}\\\cline{2-10}
& {\tiny 0.999} & {\tiny 9.31} & {\tiny 2.00} & {\tiny 27.9} &
{\tiny -2.01} & {\tiny 9.31} & {\tiny 2.00} & {\tiny 14.0} &
{\tiny 0}\\\hline
& {\tiny 0.3} & {\tiny 3.23} & {\tiny -2.18(2.73)} &
{\tiny 3.65} & {\tiny -6.20(2.27)} & {\tiny 3.23} & {\tiny -2.18(2.27)}
& {\tiny 3.43} & {\tiny -4.09(2.27)}\\\cline{2-10}
{\tiny 0} & {\tiny $\Omega_{sep}$} & {\tiny 14.0} & {\tiny 0} & {\tiny $\infty$} &
{\tiny -4.00} & {\tiny 14.0} & {\tiny 0} & {\tiny 27.0} &
{\tiny -1.99}\\\cline{2-10}
& {\tiny 0.999} & {\tiny 9.31} & {\tiny 2.00} & {\tiny 27.9} &
{\tiny -2.01} & {\tiny 9.31} & {\tiny 2.00} & {\tiny 14.0} &
{\tiny 0}\\\hline
& {\tiny 0.3} & {\tiny 2.74} & {\tiny -2.73(2.74)} &
{\tiny 3.9} & {\tiny --6.46(2.27)} & {\tiny 3.23} & {\tiny -2.46(2.27)}
& {\tiny 2.89} & {\tiny -4.75(2.22)}\\\cline{2-10}
{\tiny +1} & {\tiny $\Omega_{sep}$} & {\tiny 14.0} & {\tiny 0} & {\tiny $\infty$} &
{\tiny -4.00} & {\tiny 14.0} & {\tiny 0} & {\tiny 26.6} &
{\tiny -1.99}\\\cline{2-10}
& {\tiny 0.999} & {\tiny 9.31} & {\tiny 2.01} & {\tiny 28.0} &
{\tiny -2.01} & {\tiny 9.31} & {\tiny 2.00} & {\tiny 14.0} &
{\tiny 0}\\\hline
\end{tabular}
	\caption{Age of the Universe, deceleration parameter and maximal redshift (for $\Omega_m=0.3$) in models I-IV. $\Omega_{sep}$ is equal to 0.6345 when $k=-1$ for models I and IV and it is equal to 0.7022 when $k=+1$ for the same models. For all the rest cases it is $\Omega_{sep}=2/3$.}
	\label{table}
\end{table*}

One can see from tab. \ref{table} that the age is very different and is acceptable for most cases comparing to existing bound given by the age of globular clusters, in agreement with cosmic nucleo-chronology \cite{Kra03}
\begin{eqnarray}
	t_{obs}=11.2 \, 10^9 \,\textrm{Yrs}.
	\label{tobs}
\end{eqnarray}

In particular, for models I and III the age is 14 GYrs quite independently on k for $\Omega_m=\Omega_{sep}$. The age is larger than (\ref{tobs}) also for models II and IV for $\Omega_m\geq\Omega_{sep}$. Notice also infinite age for separatrix in model II which follows from (\ref{sol2sep}). No one of the models with $\Omega_m=0.3$ pass the age constraint (\ref{tobs}).

Deceleration parameter characterizes the second derivative of the scale factor at present
\begin{eqnarray}
	q_0=-\frac{\ddot a(t_0)}{\dot a^2(t_0)}a(t_0).
\end{eqnarray}
Observational bounds on it can be found in \cite{Riess} with
\begin{eqnarray}
	-1.35\leq q_0 \leq -0.15,
	\label{q0obs}
\end{eqnarray}
within $3\sigma$ CL.

In tab. \ref{table} we provide values for this parameter in all four models under consideration. Notice that solutions for separatrix are ``ballistic'' with $q_0=0$ and age 14 GYrs for models I and III and also the same for model IV with $\Omega_m=1$. For models with de-Sitter asymptotic such as II and IV $q_0$ is quite large and negative. Deceleration parameter comes to the region (\ref{q0obs}) only for model IV with $\Omega_{sep}\leq\Omega_m\leq 1$ with any spatial curvature parameter.

In all four models the maximal redshift for $\Omega_m=0.3$ is computed from
\begin{eqnarray}
	1+z=\frac{a_0}{a},
	\label{redshift}
\end{eqnarray}
and we find it $z_{max}\approx 2.3$ for models II-IV and a little bit higher for model I.

Putting everything together, it is clear that model IV with $\Omega_m>\Omega_{sep}$ passes all tests up to now.

\section{Supernovae and radio galaxies fit}

In what follows we use supernovae and radio galaxies data to put additional constraints on GX models. For this purpose we use the same data as in \cite{Daly}. It consist of 71 supernovae from the Supernova Legacy Survey \cite{Astir}, 157 ``Gold'' supernovae of \cite{Riess} and 20 radio galaxies of \cite{Guerra}. So in total we have 248 sources with redshifts between zero and 1.8.
\begin{table*}[htp]
	\centering
\begin{tabular}{|c|c|c|c|c|c|c|c|c|c|}
\hline
& \multicolumn{3}{|c|}{\footnotesize I} & \multicolumn{3}{|c|}{\footnotesize %
III} & \multicolumn{3}{|c|}{\footnotesize IV} \\ \cline{2-10}
{\footnotesize $\Omega_{m}$} & {\footnotesize k=1} & {\footnotesize k=0} & 
{\footnotesize k=-1} & {\footnotesize k=1} & {\footnotesize k=0} & 
{\footnotesize k=-1} & {\footnotesize k=1} & {\footnotesize k=0} & 
{\footnotesize k=-1} \\ \hline
{\footnotesize 0,6} & {\footnotesize 1,314} & {\footnotesize 1,496} & 
{\footnotesize 1,668} & {\footnotesize 1,478} & {\footnotesize 1,496} & 
{\footnotesize 1,513} &  &  &  \\ \hline
{\footnotesize 0,635} & {\footnotesize 1,459} & {\footnotesize 1,669} & 
{\footnotesize 1,846} & {\footnotesize 1,661} & {\footnotesize 1,669} & 
{\footnotesize 1,676} & {\footnotesize 4,798} & {\footnotesize 3,933} & 
{\footnotesize 3,423} \\ \hline
{\footnotesize 0,65} & {\footnotesize 1,535} & {\footnotesize 1,750} & 
{\footnotesize 1,927} & {\footnotesize 1,747} & {\footnotesize 1,750} & 
{\footnotesize 1,753} & {\footnotesize 4,436} & {\footnotesize 3,681} & 
{\footnotesize 3,233} \\ \hline
{\footnotesize 0,67} & {\footnotesize 1,628} & {\footnotesize 1,846} & 
{\footnotesize 2,020} & {\footnotesize 1,846} & {\footnotesize 1,844} & 
{\footnotesize 1,843} & {\footnotesize 4,076} & {\footnotesize 3,427} & 
{\footnotesize 3,036} \\ \hline
{\footnotesize 0,703} & {\footnotesize 1,851} & {\footnotesize 2,065} & 
{\footnotesize 2,230} & {\footnotesize 2,077} & {\footnotesize 2,065} & 
{\footnotesize 2,054} & {\footnotesize 3,412} & {\footnotesize 2,948} & 
{\footnotesize 2,658} \\ \hline
{\footnotesize 0,8} & {\footnotesize 2,547} & {\footnotesize 2,711} & 
{\footnotesize 2,833} & {\footnotesize 2,743} & {\footnotesize 2,711} & 
{\footnotesize 2,683} & {\footnotesize 2,227} & {\footnotesize 2,050} & 
{\footnotesize 1,934} \\ \hline
{\footnotesize 0,9} & {\footnotesize 3,341} & {\footnotesize 3,428} & 
{\footnotesize 3,491} & {\footnotesize 3,455} & {\footnotesize 3,428} & 
{\footnotesize 3,402} & {\footnotesize 1,571} & {\footnotesize 1,525} & 
{\footnotesize 1,494} \\ \hline
{\footnotesize 0,99} & {\footnotesize 4,080} & {\footnotesize 4,090} & 
{\footnotesize 4,096} & {\footnotesize 4,093} & {\footnotesize 4,090} & 
{\footnotesize 4,085} & {\footnotesize 1,284} & {\footnotesize 1,282} & 
{\footnotesize 1,281} \\ \hline
\end{tabular}
	\caption{Results of the fit.}
	\label{tab2}
\end{table*}
\begin{figure}[hbp]
	\begin{center}
		\includegraphics[width=5in]{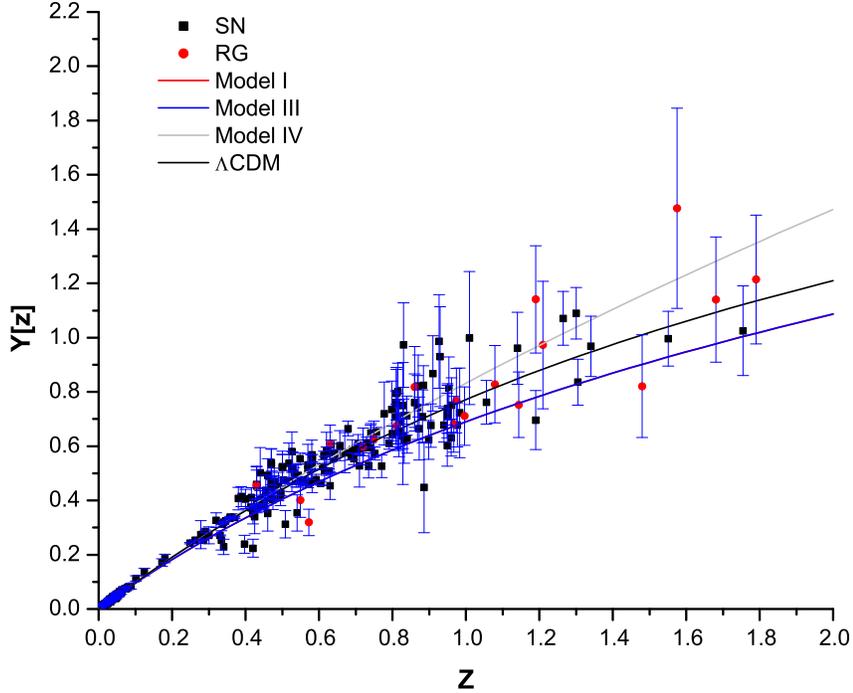}
	\end{center}
	\caption{Best fit curves for models I,III and IV. The curves for all curvatures, $k=0, +1, -1$, do coincide.}
	\label{fig1}
\end{figure}
Dimensionless coordinate distance defined as usual (see, e.g. \cite{Peebles})
\begin{eqnarray}
	Y(a)=\frac{H_{0}a_{0}}{c}\int _t^{t_0}\frac{c(t)dt}{a(t)},
\end{eqnarray}
and it can be expressed in a more convenient form as $Y(z)$ with the help of (\ref{redshift}).

First of all, for standard $\Lambda$CDM model with $\Omega_m=0.3$, $\Omega_\Lambda=0.7$ we find $\chi^2=1.16$ with 245 Dof. Then for each model cosmological equations are solved numerically and dimensionless coordinate distance $Y(z)$ is calculated. The fit was obtained by the usual least square technique for all spatial curvatures with $0.6<\Omega_m<0.999$ for all four models. Results of the fit for models I,III,IV are shown in tab. \ref{tab2}. The model II does not fit the data well and results for it are not included in the table.

As one can see from tab. \ref{tab2} for model IV best fit is obtained when $\Omega_{m}=0.999$ and it is the same for different values of $k$. In models I,III the fit is better with smaller $\Omega_{m}$. It is interesting to note that curves for models I and III with different $k$ practically coincide with each other. This is due to the fact that in the cosmological equations the corresponding term is $1-\frac{2k}{3\pi^2}$, and is close to 1 with all curvatures. There is also another symmetry between model I,III. As one can see from tab. \ref{tab2} in case when $k=0$ solutions for $Y(z)$ are the same for both models. Clearly, the best fit among models is given by model IV with high $\Omega_m$.

In fig. \ref{fig1} we show our results for models I, III, IV along with with 248 sources, including 228 supernovae and 20 radio galaxies. Since the character of solutions changes for $\Omega_m<\Omega_{sep}$ for models I,III we took $\Omega_{sep}$ as a best fit. As already mentioned before, for models I,III our curves are the same so in fig. \ref{fig1} they are shown as the same line.

\section{Conclusions}

Analytic solutions are obtained for all Gurzadyan-Xue models. Solutions for models II and IV are in explicit form. The most simple ones turn out the solutions for the separatrix $\Omega_{sep}$ for each model: power law for models I, III and IV and exponential one for model II.

The presence of the separatrix, defining two families of solutions, appears equally crucial at comparison with observational data, for example, leading to the preference of higher matter density $\Omega_m\geq\Omega_{sep}$ in all models. This is mainly, because the maximal redshift in models with $\Omega_m<\Omega_{sep}$ is small, although the age can be acceptable, and on the contrary, low matter density models have shorter age. Note, however, that rather simple matter density models have been considered, mainly, to probe the value of the separatrix.  

Thus, the analysis confirmed the crucial role of the separatrix not only for the behavior of the solutions, but also at comparison with observations, which a priori is never trivial since the initial equations are quite different.

\end{document}